\documentclass[aip, amsmath, amssymb, reprint, 9pt]{revtex4-1}
\usepackage{dcolumn, bm, indentfirst}
\usepackage[title]{appendix}
\usepackage{graphicx, tikz, orcidlink}
\usepackage{epstopdf}
\usepackage{ascmac}
\usepackage{ulem}
\usepackage{cancel}
\usepackage{here}

\usepackage[]{hyperref}
\usepackage{xcolor}
\definecolor{AIPBlue}{RGB}{61, 180, 229}
\hypersetup{
colorlinks,
linkcolor=AIPBlue,
citecolor=AIPBlue,
urlcolor=AIPBlue
}

\begin{document}
\title{Numerically ``exact'' simulations of a quantum Carnot cycle: Analysis using thermodynamic work diagrams}
\date{Last updated: \today}

\author{Shoki Koyanagi \orcidlink{0000-0002-8607-1699}}
\author{Yoshitaka Tanimura \orcidlink{0000-0002-7913-054X}}
\email[Author to whom correspondence should be addressed: ]{tanimura.yoshitaka.5w@kyoto-u.jp}
\affiliation{Department of Chemistry, Graduate School of Science,
Kyoto University, Kyoto 606-8502, Japan}

\begin{abstract}
We investigate the efficiency of a quantum Carnot engine based on open quantum dynamics theory. The model includes time-dependent external fields for the subsystems controlling the isothermal and isentropic processes and for the system--bath (SB) interactions controlling the transition between these processes. Numerical simulations are conducted in a nonperturbative and non-Markovian SB coupling regime using the hierarchical equations of motion under these fields at different cycle frequencies. The work applied to the total system and the heat exchanged with the baths are rigorously evaluated. In addition, by regarding quasi-static work as free energy, we compute the quantum thermodynamic variables and analyze the simulation results using thermodynamic work diagrams for the first time. Analysis of these diagrams indicates that, in the strong SB coupling region, the fields for the SB interactions are major sources of work, while in other regions, the field for the subsystem is a source of work. We find that the maximum efficiency is achieved in the quasi-static case and is determined solely by the bath temperatures, regardless of the SB coupling strength, which is a numerical manifestation of Carnot’s theorem. 
\end{abstract}
\maketitle

\section{Introduction}
\label{sec.intro}
In 1824, Carnot proposed a reversible heat engine (subsequently named the Carnot engine) that operates between a high-temperature (hot) bath at temperature $T_1$ and a low-temperature (cold) bath at temperature $T_2$, with gas used as a working medium.\cite{Carnot1824} This heat engine performs the Carnot cycle, which consists of four reversible processes: (i)~isothermal expansion, (ii)~isentropic expansion, (iii)~isothermal compression, and (iv)~isentropic compression. Here, heat $Q_1$ is absorbed from the hot bath, and work is produced during the isothermal expansion process; heat $Q_2$ is expelled to the cold bath, and some work is lost in the isothermal compression process. The isentropic processes bridge the two isothermal processes quasi-statically by lowering or raising the temperature of the system to $T_2$ or $T_1$, respectively. The work is done while adiabatically reducing the temperature, and the net heat gain is equal to the work done, i.e., $W=Q_1-Q_2$. Carnot proved that (a)~the thermal efficiency of this engine is the maximum that is possible and that (b)~this maximum efficiency is determined solely by the temperatures of the heat baths (Carnot’s theorem). Moreover, in the Carnot cycle, the relationship $Q_1/T_1 -Q_2/T_2=0$ is satisfied. To elucidate the characteristic features of the Carnot cycle, a thermodynamic work diagram (the $P$--$V$ diagram) was later introduced by Clapeyron.\cite{Clapeyron1834} The work done by the system is represented by the area enclosed by the curves in the $P$--$V$ diagram.

In 1848, Thomson (later Lord Kelvin) established that the temperatures appearing in the above formula should be regarded as thermodynamic (or absolute) temperatures.\cite{Thomson1848} In the 1850s, the foundations of the second law of thermodynamics were established by Clausius on the basis of Carnot’s result.\cite{Clausius1850,Clausius1856} This result was generalized by Thomson and Clausius as $\oint dQ/T =0$, where $T$ is the temperature of the heat source and $Q$ is the heat reversibly transferred to the system. In 1865, Clausius introduced the concept of entropy, which is defined as $dS=dQ/T$, and expressed the second law of thermodynamics for heat $Q$ as $\oint d Q / T < \oint d S = 0$ in any irreversible process.\cite{Clausius1865} Entropy is an extensive variable that characterizes the reversibility or irreversibility of thermal processes based on the Carnot cycle, and its conjugate intensive variable is temperature. The Carnot cycle was a key element in the construction of thermodynamics.

Recently, the advent of nanotechnology has led to thermodynamic investigations being extended to the quantum regime,\cite{Binder2018,e15062100, lambert2019modelling, ReichmanPhysRevB2013,SanchezPhysRevB2014,GalperinPRL2015,GalperinJCP2020,Brandao3275,PhysRevB.97.085435,PhysRevLett.120.120602,PhysRevE.98.012113,PhysRevLett.124.160601,PhysRevBUncertenty2020,RevModPhys.92.041002, PhysRevA.85.063811, PhysRevB.86.014501, PhysRevApplied.17.064022, PhysRevB.100.035407, PhysRevLett.127.060601, PhysRevLett.127.150401,PRXQuantum.3.020305} in which the work and heat to be manipulated are quantized.\cite{JarEXE2015,AdiaEXE2016,Martinez2016,KlassRMP2017,LundJosefsson2018,ReimannPRL2018,LundJosefsson2019,nanolettDomenic2019,HuiPRL2020,DEOLIVEIRA2020,Hern_ndez_G_mez_2021,SanchezdNatureCarnot2022} However, such studies involve fundamental difficulties because the time evolution of the main system (subsystem) is described by quantum mechanics, while the thermal effects of the system are described by macroscopic thermodynamics and statistical mechanics for the equilibrium state. For example, in an investigation of isothermal processes, we must explicitly treat the system--bath (SB) interactions so as to maintain the subsystem in a state of thermal equilibrium under the influence of an external force. Because the energy scale of the SB interactions is comparable to the energy of the subsystem, we must consider not only the subsystem, but also the SB interactions in a quantum-mechanically consistent manner. Moreover, the quantum description of the heat bath is significant because the subsystem and the bath are entangled (bath entanglement), particularly in the low-temperature case.\cite{T06JPSJ,T20JCP,T14JCP,T15JCP} In an adiabatic process, a thermodynamic system with a large number of degrees of freedom is usually assumed to be in an equilibrium state with a quasi-static change in an extensive thermodynamic variable. An isolated quantum subsystem, however, cannot reach a thermal equilibrium state of its own, and thus its temperature cannot be uniquely defined during adiabatic processes. Additionally, the definitions of thermodynamic variables in a small subsystem, in particular extensive variables such as magnetization and strain, are not clear in the quantum case.

In the present paper, to clarify the relationship between quantum mechanics and statistical thermodynamics, we describe the results of quantum simulations of a Carnot cycle on the basis of an SB model. Although such investigations have been conducted in the framework of open quantum dynamics theories, even for strong SB coupling cases (e.g., for heat transport,\cite{doi:10.1063/1.1603211,PhysRevLett.94.034301,PhysRevB.73.205415,doi:10.1021/jp5091685,doi:10.1080/00018730802538522,VELIZHANIN2008325,RevModPhys.83.131,2015JCao,PhysRevLett.111.214301,PhysRevB.95.064308,PhysRevE.99.042130,doi:10.1063/1.5084949,PhysRevB.103.104304,KT15JCP, GalperinPhysRevB2015} heat engines and refrigerators,\cite{PhysRevE.76.031105,PhysRevE.89.032115,2011Otto,doi:10.1146/annurev-physchem-040513-103724,PhysRevX.5.031044,KT16JCP,PhysRevE.97.022130,2018Markov,PhysRevE.98.012117,2020Otto,PhysRevA.102.012217,e23091149, OttoNori2020, Non-MarkovWiedmann_Ankerhold2020,OttoWiedmann_Ankerhold2021,Xu_Ankerhold2022,TensorNetPhysRevX2020,PhysRev2020SBcouple,PhysRevEOtto2020,Stirling2021,Segal2022PRE} entropy production,\cite{GOYAL2020122627,GalperinPhysRevB2021,ST20JCP} nonequilibrium cases,\cite{PhysRevB.91.224303,PhysRevBUncertenty2020} and quantum information problems,\cite{2016Skrzypczyk,PhysRevX.7.021003} in addition to being applied as a context for the Jarzynski equality,\cite{Jarzynski2004,doi:10.1143/JPSJ.69.2367,ST21JPSJ} fluctuation theorems,\cite{PhysRevLett.102.210401,RevModPhys.81.1665,2012,RevModPhys.83.771,PhysRevE.101.052116,WhitneyPhysRevB2018} and Maxwell's demon\cite{PhysRevLett.97.180402,RevModPhys.81.1,leff2002maxwell}), fully quantum investigations face difficulties because of the lack of a consistent thermodynamic formulation for a thermal system described by a Hamiltonian with external time-dependent perturbations.

Recently, it was shown that heat can be defined as the change in the heat-bath energy.\cite{KT15JCP,KT16JCP} The key to investigating this problem is the hierarchical equations of motion (HEOM) formalism, which enables the evaluation of the internal energies of not only the subsystem, but also the bath and the SB interactions, even in low-temperature, non-Markovian, and nonperturbative conditions.\cite{TK89JPSJ1,T90PRA,IT05JPSJ,T06JPSJ,T14JCP,T15JCP,T20JCP}
We show that, by introducing a time-dependent SB interaction, thermal transitions between isothermal and adiabatic states can also be investigated. By using HEOM to calculate the change in work and bath energy applied to the entire system, the efficiency of the heat engine under an arbitrary time-dependent external field can then be evaluated without assumptions.\cite{KT22JCP1}

To enable thermodynamic analysis, the thermodynamic variables can be treated based on the SB model. For this purpose, we adopt the minimal work principle for a total isolated system, expressed as $W_{\rm tot}(t) \ge \Delta F_{\rm tot}(t)$, where $W_{\rm tot} ( t )$ is the work done by external fields and $\Delta F_{\rm tot}(t)$ is the change of free energy, and define the ``quasi-equilibrium'' Helmholtz energy (qHE) as $\Delta F_{\rm tot} (t) = W_{\rm tot}^{\rm qst}(t)$ with the system driven quasi-statically by external fields. Although $W_{\rm tot} ( t )$ cannot be evaluated within the framework of regular open quantum dynamics theories (because the number of degrees of freedom of the bath has been reduced), we can evaluate this quantity indirectly using the hierarchical elements in the HEOM formalism. This is because, in the HEOM formalism, the higher hierarchical elements store information about the higher cumulant of the bath coordinates, as previously demonstrated.\cite{KT15JCP,KT16JCP,PhysRevB.95.064308,ST20JCP,ST21JPSJ} 
We then show that the Kelvin--Planck statement (or heat engine statement) of the second law of thermodynamics is only validated when we introduce a time-dependent SB interaction that describes energy conservation for adiabatic transitions.\cite{KT22JCP1} 

To this end, we extend previous research to verify Carnot's theorem using the HEOM approach. The Carnot engine is described as a two-level subsystem (A) coupled with harmonic heat baths (B). To control the isothermal process, we consider the external field for the system (the isothermal driving field) and, to turn the hot and cold baths on and off, we introduce two time-dependent SB interactions (the adiabatic transition fields). We then investigate the thermodynamic efficiency of the Carnot cycle under various physical conditions. 
In addition, using the qHE formalism, thermodynamic work diagrams of external forces (such as stresses) and their conjugate variables (such as strains), similar to Clapeyron’s $P$--$V$ diagram,\cite{Clapeyron1834} are introduced to analyze the work done in the system. This extension is useful for analyzing experimental results in the quantum regime, where the quantized work and heat are to be manipulated.\cite{JarEXE2015,AdiaEXE2016,KlassRMP2017, LundJosefsson2018,ReimannPRL2018,LundJosefsson2019,nanolettDomenic2019,HuiPRL2020,DEOLIVEIRA2020,Hern_ndez_G_mez_2021}

The remainder of this paper is organized as follows. In Sec.~\ref{sec:model}, we introduce the SB Hamiltonian and the HEOM formalism. We then describe the scheme used to calculate various thermodynamic variables based on open quantum dynamics theory. In Sec.~\ref{QuantumCarnot}, we explain our model of the quantum Carnot cycle and present simulation results under the isothermal driving and adiabatic transition fields. The conjugate properties of these variables are calculated by regarding quasi-static work as free energy. Thermodynamic work diagrams are presented as functions of these variables. Finally, Sec.~\ref{sec:conclude} presents concluding remarks.

\section{Theory}
\label{sec:model}
\subsection{Model}
We consider a subsystem A coupled to two heat baths at high and low inverse temperatures $\beta_1 = 1/k_B T_1$ and $\beta_2 =1 / k_B T_2$ as the heat sources, where $k_B$ is the Boltzmann constant. The total Hamiltonian is expressed as
\begin{align}
\hat{H}_{\rm tot}(t) =\hat{H}_{\rm A}(t) + \sum_{k=1}^2 \left( \hat{H}_{\rm I}^{k} (t) + \hat{H}_{\rm B}^{k} \right),
\label{eq:Htotal}
\end{align}
where $\hat{H}_{\rm A}(t)$, $\hat{H}_{\rm I}^{k} (t)$, and $\hat{H}_{\rm B}^{k}$ are the Hamiltonians of the system, $k$th SB interaction, and $k$th bath, respectively.
We consider a two-level system (TLS) defined as
\begin{equation}
\hat{H}_{\rm A} ( t ) = - B( t ) \hat{\sigma}_z + E \hat{\sigma}_x ,
\end{equation}
where $B(t)$ is the isothermal driving field (IDF), $E$ is the off-diagonal coupling parameter, and $\hat{\sigma}_{\alpha}$ ($\alpha = x$, $y$, or $z$) are Pauli matrices. In the case of a spin system, $B(t)$ corresponds to the longitudinal magnetic field and $E$ is the transverse electric (Stark) field. The Hamiltonian representing the $k$th SB interaction and the $k$th bath are given by\cite{KT22JCP1}
\begin{align}
\hat{H}_{\rm I}^{k}(t) = A_k (t) \hat{V}_k \sum_j c_{j}^k \left[ \hat{b}_{j}^k + (\hat{b}_{j}^k)^\dagger \right]
\end{align}
and
\begin{align}
\hat{H}_{\rm B}^{k} = \sum_j \hbar \omega_{j}^k \left[ (\hat{b}_{j}^k)^\dagger \hat{b}_{j}^k + \frac{1}{2} \right],
\end{align}
respectively, where $\hat{V}_k$ is the system operator that describes the coupling to the $k$th bath and $A_k(t)$ is the adiabatic transition field (ATF), which is introduced to describe the operation of an adiabatic wall between the system and the $k$th heat bath (e.g., the insertion or removal of the adiabatic wall or attaching or detaching the quantum system to/from the bath). Here, $\omega_{j}^k$, $c_{j}^k$, $\hat{b}_{j}^k$, and $(\hat{b}_{j}^k )^\dagger$ are the frequency, coupling strength, and annihilation and creation operators for the $j$th mode of the $k$th bath, respectively.

Due to the bosonic nature of the baths, all bath effects on the system are determined by the $k$th bath correlation function, $C_k(t) \equiv \langle \hat{X}_k(t) \hat{X}_k(0) \rangle_{\rm B}$, where $\hat{X}_k \equiv \sum_j c_{j}^k[\hat{b}_{j}^k + (\hat{b}_{j}^k) ^\dagger ]$ is the collective coordinate of the $k$th bath and $\langle \cdots \rangle_{\rm B}$ represents the average taken with respect to the canonical density operator of the baths. The bath correlation function is expressed as
\begin{align}
C_k(t)
= \int_0^\infty d\omega \, \frac{ J_k(\omega) }{ \pi }
\left[ \coth\left( \frac{\beta_k\hbar\omega}{2} \right) \cos(\omega t)
- i \sin(\omega t) \right],
\label{eq:BCF}
\end{align}
where $J_k(\omega) \equiv \pi \sum_j (c_{j}^k)^2 \delta(\omega - \omega_{j}^k)$ is the bath spectral density. The real part of Eq.~(\ref{eq:BCF}) is analogous to the classical correlation function of the bath and corresponds to the fluctuations, while the imaginary part corresponds to the dissipation. The fluctuation term is related to the dissipation term through the quantum version of the fluctuation--dissipation theorem. \cite{TK89JPSJ1,T06JPSJ} In this paper, we use the Drude spectral density function, described as
\begin{equation}
J_k ( \omega ) = \frac{\hbar \gamma_k^2 \omega}{\gamma_k^2 + \omega^2} ,
\end{equation}
where $\gamma_k$ is the inverse noise correlation of the $k$th bath.

\subsection{Hierarchical equations of motion}
In the HEOM formalism, the set of equations of motion consists of the auxiliary density operators (ADOs).\cite{TK89JPSJ1,T90PRA,IT05JPSJ, T06JPSJ,T14JCP,T15JCP,T20JCP,KT22JCP1} Here, we consider the case in which the bath correlation function, Eq.~\eqref{eq:BCF}, is written as a linear combination of exponential functions, $C_k(t) = \sum_{l=0}^{K_k} \zeta_{l}^k e^{-\nu_{l}^k |t|}$, where $\nu_l^k$, $\zeta_l^k$, and $K_k$ are the frequency, strength, and cutoff integer value for the $k$th bath obtained from a Pad{\'e} spectral decomposition scheme to reduce the hierarchy size.\cite{hu2010communication} The ADOs introduced in the HEOM are defined by $\hat{\rho}_{\vec{n}} ( t )$ with a set of indices $\vec{n} = ( n_{1}^0, \ldots, n_{1}^{K_1}, n_{2}^0, \ldots, n_{2}^{K_2})$, where $n_{l}^{k}$ represents an integer value of zero or above. The zeroth ADO, $\hat{\rho}_{\vec{0}}(t)$
with $\vec{0} = ( 0, 0, \ldots, 0)$, corresponds to the actual reduced density operator. The HEOM for the IDF and ATFs are then expressed as\cite{KT22JCP1}
\begin{align}
\frac{\partial}{\partial t} \hat{\rho}_{\vec{n}} ( t ) &= \left( - \frac{i}{\hbar} \hat{H}_{\rm A}^\times ( t ) -\sum_{k=1}^2 \sum_{l=0}^{K_k} n_{l}^k \nu_{l}^k \right) \hat{\rho}_{\vec{n}} ( t ) \nonumber \\
& - \frac{i}{\hbar} \sum_{k=1}^2 {A_k( t )} \sum_{l = 0}^{K_k} n_{l}^k \hat{\Theta}_{l}^k \hat{\rho}_{\vec{n} - \vec{e}_l^k} ( t ) \nonumber \\
& - \frac{i}{\hbar} \sum_{k=1}^2 {A_k( t )} \sum_{l = 0}^{K_k} \hat{V}_k^\times \hat{\rho}_{\vec{n} + \vec{e}_l^k} ( t ),
\label{ModelHEOM}
\end{align}
where $\vec{e}_{l}^{~k}$ is the ($K_k$+1)-dimensional unit vector. We introduce a set of fluctuation--dissipation operators as
\begin{equation}
\hat{\Theta}_{0}^k \equiv \left( \frac{ \gamma_k}{\beta_k} + \sum_{m = 1}^{K_m} \frac{ \zeta_m \gamma_k^2}{\beta_k} \frac{ 2 \gamma_k}{\gamma_k^2 - \nu_m^2} \right) \hat{V}_k^\times - \frac{i \hbar \gamma_k^2}{2} \hat{V}_k^\circ
\end{equation}
and
\begin{align}
\hat{\Theta}_{l}^k \equiv - \frac{ \zeta_l \gamma_k^2}{\beta_k} \frac{2 \nu_l}{\gamma_k^2 - \nu_l^2} \hat{V}_k^\times ,
\end{align}
where $\hat{\mathcal{O}}^\times \hat{\mathcal{P}} = [ \hat{\mathcal{O}} , \hat{\mathcal{P}} ]$ and $\hat{\mathcal{O}}^\circ \hat{\mathcal{P}} = \{ \hat{\mathcal{O}} , \hat{\mathcal{P}} \}$ for arbitrary operators $\hat{\mathcal{O}}$ and $\hat{\mathcal{P}}$, and $\nu_l$ and $\zeta_l$ are the frequency and strength, respectively.

As the temporal initial conditions, we consider the factorized initial case
$$
\hat{\rho}_{\rm tot}(0) = \hat{\rho}_{\rm A}(0) \prod_{k=1}^2 \frac{e^{- \beta_k \hat{H}_{\rm B}^{k} } }{ {\rm tr}_{\rm B}\{ e^{ - \beta_k \hat{H}_{\rm B}^{k} }\} } ,
$$
where $\hat{\rho}_{\rm A}(t)=\hat{\rho}_{\vec{0}}(t)$ is the reduced density operator of the subsystem. To obtain the bath-entangled steady-state solution described as $\hat{\rho}_{\vec{n}}(t)$, we integrate the HEOM under the periodical external fields until all of the hierarchy elements reach a steady state.

Using the zeroth member of the hierarchy $\hat{\rho}_{\vec{0}}(t)$, the nonequilibrium internal energy of the system (or the expectation value of the system energy) at time $t$ is evaluated as
\begin{align}
U_{\rm A}^{\rm neq} (t) = {\rm tr}_{\rm A} \left\{ \hat{H}_{\rm A} ( t ) \hat{\rho}_{\vec{0}} ( t ) \right\}.
\label{expectHA}
\end{align}
The nonequilibrium internal energy of the $k$th SB interaction is then expressed as\cite{KT15JCP,KT16JCP,ST20JCP,ST21JPSJ}
\begin{align}
U_{{\rm I}_k}^{\rm neq} (t)= A_k ( t ) \sum_{l = 0}^{K_k} {\rm tr}_{\rm A} \left\{ \hat{V}_k \hat{\rho}_{\vec{e}_l ^k} ( t ) \right\},
\label{expectHI}
\end{align}
where $\vec{e}_l$ is the index for the first-order hierarchical member.

Using the HEOM formalism, we can evaluate the $k$th bath energy as\cite{ST20JCP,KT22JCP1}
\begin{align}
\frac{\partial}{\partial t} U_{{\rm B}_k}^{\rm neq} (\beta_k ; t)&= A_k ( t ) \sum_{l = 0}^{K_k} \nu_{l} {\rm tr} \{ \hat{V} \hat{\rho}_{\vec{e}_l} ( t ) \} \nonumber \\
& + A_k^2 ( t ) \gamma^2 {\rm tr}_{\rm A} \left\{ \hat V^2 \hat{\rho}_{\vec{0}} ( t ) \right\}.
\label{expectHB}
\end{align}
Because there is no external force applied to the bath, the $k$th bath heat can be obtained by integrating the above equation as
\begin{align}
Q_{{\rm B}_k} (\beta_k; t) = \Delta U_{{\rm B}_k}^{\rm neq}(\beta_k; t)
\label{U_Brd2}
\end{align}
with $\Delta U_{{\rm B}_k}^{\rm neq} (\beta_k; t) \equiv U_{{\rm B}_k}^{\rm neq} (\beta_k; t+\delta t)- U_{{\rm B}_k}^{\rm neq} (\beta_k; t)$, where $\delta t$ is the time step of the thermal process.

In this study, work is defined as the change in energy from one state to another state under a time-dependent perturbation expressed as
\begin{align}
W_{\rm tot} (t) = \int_{t}^{t+\delta t} dt' {\rm{tr}}\left\{ \frac{d \hat H_{\rm tot} (t')}{dt'} \hat \rho_{\rm tot}(t') \right\}.
\label{work}
\end{align}
Because there is no external field for the bath, we have $W_{\rm tot} (t) =W_{{\rm A}+{\rm I}} (t )$, where
\begin{equation}
W_{{\rm A}+{\rm I}} (t ) \equiv W_{\rm A} ( t) + W_{{\rm I}_1} (t ) + W_{\rm I_2} (t ).
\label{eq:W}
\end{equation}
Using the HEOM, each component is evaluated as
\begin{align}
W_{\rm A} (t )= \int_{t}^{t+\delta t} dt' {\rm tr}_{\rm A} \left\{ \frac{\partial {\hat H}_{\rm A}( t' )}{\partial t} \hat{\rho}_{\vec{0}}(t') \right\}
\label{qHEWA}
\end{align}
and
\begin{align}
W_{{\rm I}_k} ( t ) = \int_{t}^{t+\delta t} dt'\frac{d A_k (t')}{d {t'}} \sum_{l = 0}^{K_k} {\rm tr}_{\rm A} \{ \hat{V}_k \hat{\rho}_{\vec{e}_l^k} ( t' ) \}.
\label{qHEWI}
\end{align}
By this definition, when $W_{\rm tot} ( t ) > 0$, work is done from the outside to the total system and is called ``positive work.'' 

Using the above, we can evaluate the $k$th bath energy for $A_k ( t ) > 0$ more accurately than from Eq.~\eqref{expectHB} as
\begin{align}
\frac{\partial}{\partial t} U^{\rm neq}_{\rm B_k} (\beta_k; t )& = 
- \frac{\partial}{\partial t} \left[ U^{\rm neq}_{\rm A}(t) 
+ U^{\rm neq}_{\rm I_k}(t) \right] \nonumber \\
&+ \frac{\partial}{\partial t} \left[ W_{\rm A}(t) + W_{\rm I_k}(t) \right].
\end{align}

\subsection{Quasi-equilibrium Helmholtz energy and thermodynamic variables}
\label{NEFE}

Using the above definition of work with the HEOM, the efficiency of the system driven by the IDF and ATFs can be evaluated numerically and rigorously for any cycle speed. Nevertheless, we concentrate our discussion to the quasi-static case and attempt to quantify the quantum thermodynamic variables by comparing the physical quantities calculated in the HEOM with the thermodynamic quantities evaluated in the qHE, as explained below.

The work for the total system satisfies the minimum work principle, expressed as\cite{Lenard1978,tasaki2000,Oono2017}
\begin{align}
W_{\rm tot} (\beta; t) \ge \Delta F_{{\rm tot}} (\beta; t),
\label{Ineq}
\end{align}
which corresponds to the second law of thermodynamics. 
For the SB model, the above inequality has been verified numerically by placing the real-time HEOM for work on the left-hand side and the imaginary-time HEOM for the free energy on the right-hand side.\cite{T14JCP,T15JCP,ZT22JCP} This is because, even when ${\hat H}_{A}$ and ${\hat H}_{I}$ are weakly dependent on time, the steady-state solution of the HEOM is represented as the reduced density operator of the total equilibrium state, expressed as ${\rm tr}_{\rm B}\{\exp[-\beta ( {\hat H}_{\rm A}+{\hat H}_{\rm I}+{\hat H}_{\rm B}) ]\}$.\cite{T14JCP,T15JCP} It is known that, for a quasi-static process, the equality can be expressed within the level of numerical accuracy as\cite{ST20JCP,ST21JPSJ,KT22JCP1}
\begin{align}
\Delta F_{{\rm tot}}^{\rm qst}(\beta; t) = W_{\rm tot}^{\rm qst} (\beta; t).
\label{Fneq}
\end{align}
While the above equation holds for the total system, rather than the reduced system, we can utilize the qHE using the HEOM formalism because, through the use of the HEOM elements, we can evaluate the total work of the system, as illustrated in Eqs.~\eqref{work}--\eqref{qHEWI}.
For each component, we also have $\Delta F_{\alpha}^{\rm qst} (\beta; t ) = W_{\alpha}^{\rm qst} (\beta; t)$ for $\alpha= {\rm A}$, ${\rm I}_1$, and ${\rm I}_2$, where $W_{\alpha}^{\rm qst} (\beta; t)$ is defined by Eqs.~\eqref{qHEWA} and~\eqref{qHEWI} using the quasi-static solution of the HEOM, which is expressed as $\hat{\rho}_{\vec{n}}^{\rm qst} (t)$. \cite{KT22JCP1}

In the Carnot cycle, the subsystem interacts with a single bath at a time. For external fields, such as the magnetic field $B(t)$ and stress $A_k(t)$, we then have that $\Delta F_{\rm tot}^{\rm qst} (\beta; t ) = \Delta F_{\rm A+I}^{\rm qst} (\beta; t)$ for $\beta=\beta_k$ with $k=1$ or 2, where $\Delta F_{\rm A+I}^{\rm qst} (\beta_k; t)\equiv \Delta F_{\rm A}^{\rm qst} (\beta_k; t)+\Delta F_{\rm I_k}^{\rm qst} (\beta_k; t)$. 
Then, we can define the conjugate variables, such as the magnetization and strain, as\cite{KT22JCP1}
\begin{equation}
\label{DefnoneqM}
M^{\rm qst} (\beta_k; t) \equiv - \frac{\partial \Delta F_{\rm A}^{\rm qst} (\beta_k; t)}{\partial {B(t)}}
\end{equation}
and
\begin{equation}
\label{DefnoneqD}
D_k^{\rm qst} (\beta_k; t) \equiv - \frac{\partial \Delta F_{{\rm I}_k}^{\rm qst} (\beta_k; t)}{\partial {A_{k} (t)}}.
\end{equation}
Thus, we can evaluate the above variables in terms of the state described by the ADOs at time $t$ as
\begin{equation}
M^{\rm qst} (\beta_k; t) = {\rm tr}_{\rm A} \left\{ \hat{\sigma}_z \hat{\rho}_{\vec{0}} ( t ) \right\}
\end{equation}
and
\begin{equation}
D_k^{\rm qst} (\beta_k; t)
= - \sum_{l = 0}^{K_k} {\rm tr}_{\rm A} \left\{ \hat{V}_k \hat{\rho}_{\vec{e}^k_l} ( t ) \right\}.
\end{equation} 
The change in the quasi-equilibrium Boltzmann entropy from time $t$ to $t+\delta t$ can also be evaluated by differentiating the qHE with respect to the $k$th bath inverse temperature $\beta_k$ as\cite{ST20JCP}
\begin{align}
\label{defqBE}
\Delta S_{\rm A+I}^{\rm qst} ( \beta_k ; t ) \equiv k_{\rm B} \beta_k^2 
\frac{\partial \Delta F^{\rm qst}_{\rm A+I} ( \beta_k ; t )}{\partial \beta_k},
\end{align}
where we consider the case in which the subsystem interacts with the $k$th bath at that time. A comparison of this definition with the von Neumann entropy using the simulation results for the Carnot cycle is given in Appendix \ref{sec.TimeEntropy}. 

Note that $M^{\rm qst} (\beta; t)$, $D_k^{\rm qst} (\beta; t)$, and $\Delta S_{\rm A+I}^{\rm qst} ( \beta ; t ) $ are state variables in the quasi-static state because they are uniquely determined by the state specified by the quasi-equilibrium distribution at $t$ and are independent of the pathway of work. When the subsystem consists of $n$ noninteracting spins that are independently coupled to the heat bath, the magnitudes of the above variables are proportional to $n$. Thus, they are extensive properties, while $B(t)$, $A_k (t)$, and $\beta$ are intensive properties. 

The quasi-static isothermal process at temperature $T_k=1/k_B \beta_k $ is then expressed as
\begin{align}
d \Delta F^{\rm qst}_{\rm A+I} (\beta_k; t)&= - \Delta S_{\rm A+I}^{\rm qst} ( \beta_k ; t ) d T_k - M^{\rm qst}(\beta_k; t) d B \nonumber \\
&- D_k^{\rm qst} (\beta_k; t) d A_k.
\end{align}

\section{Quantum Carnot cycle}
\label{QuantumCarnot}

The conventional Carnot cycle consists of a system, described by pressure $P$ and volume $V$, that interacts with hot and cold baths. Although isothermal processes can be easily described by changing an external perturbation $B(t)$, there are several difficulties in realizing the Carnot cycle, even theoretically, in a nanoscale quantum system. For example, to conduct a microscopic investigation of the Carnot cycle, the thermal work done by the insertion or removal of the adiabatic wall must be included, because otherwise the energy conservation of the total system would be violated. Accordingly, because a small isolated system cannot reach thermal equilibrium by itself, an isentropic process cannot be achieved spontaneously by turning off the heat bath. Finally, although the Carnot cycle has been commonly characterized by the $P$--$V$ diagram, in a small quantum system, the definitions of extensive variables are unclear. As we explained, the qHE provides the necessary means for analyzing the simulation results of the quantum Carnot cycle obtained from the HEOM approach.

\subsection{Case under factorized assumption}
\label{factorized}
\begin{figure}[!t]
\includegraphics[width=7cm]{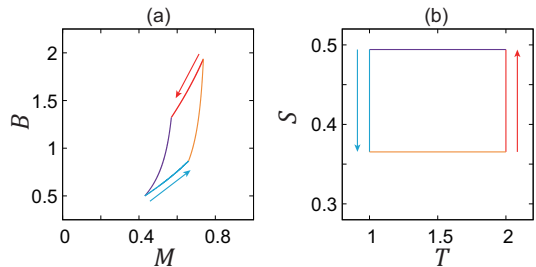}
\caption{\label{BM0} (a) $B$--$M$ diagram of Carnot engine driven by $B(t)$ with $B_0 =2$ and $E = 0.5$ evaluated under the factorized assumption, i.e., $F_{\rm A}(t) = - \ln {\rm tr}_{\rm A} \{ \exp [ -\beta H_{\rm A} (t)] \}/\beta$. Here, the blue and red curves represent the cold ($\beta_2=1.0$) isothermal expansion and hot ($\beta_1=0.5$) isothermal compression processes, while the purple and orange curves represent the hot isentropic expansion and cold isentropic compression processes, respectively. The cycle starts from the red arrow and evolves in a counterclockwise fashion over time (heat engine). (b) $T$--$S$ diagram under the same condition as (a).
}
\end{figure}
Before presenting the computational results, we will first illustrate the $B$--$M$ diagram under the assumption that the free energy of the TLS is determined by the partition function of the isolated TLS with Markovian and extremely weak SB interactions (factorized assumption). Thus, from $F_{\rm A}^0 (\beta; t) = - \ln {\rm tr} \{ \exp [ -\beta \hat{H}_{\rm A} (t)] \}/\beta$, we have
\begin{equation}
F_{\rm A}^0 (\beta; t) = - \frac{1}{\beta} \ln \left[ 2 \cosh \left( \beta \sqrt{ B^2(t) + E^2} \right) \right].
\label{FA0}
\end{equation}
The conjugate variable $M_{\rm A}^0(\beta_k; t) \equiv - {\partial F_{\rm A}^0 (\beta_k; t) }/{\partial B}$ at the inverse temperature $\beta_k$ is then evaluated as
\begin{equation}
M_{\rm A}^0 (\beta_k; t) = \frac{B(t)}{\sqrt{ B^2(t) + E^2} }\tanh \left( \beta_{k} \sqrt{ B^2(t) + E^2} \right) .
\label{M_A0}
\end{equation}
The entropy can also be evaluated by differentiating $F^0_{\rm A} ( \beta ; t )$ with respect to $\beta$ as 
\begin{equation}
\begin{split}
S^0_{\rm A} ( \beta ; t ) &= k_{\rm B} 
\left\{ \ln \left[ 2 \cosh \left( \beta \sqrt{B^2 ( t ) + E^2} \right) \right] \right. \\
& \quad \quad \left. - \beta \sqrt{B^2 ( t ) + E^2} \tanh( \beta \sqrt{B^2 ( t ) + E^2} ) \right\}.
\end{split}
\end{equation}
This value agrees with the von Neumann entropy defined by Eq.~\eqref{eq:Neumann-entropyt}, because the subsystem is isolated.

Under the factorized assumption, we can depict the $B$--$M$ diagram for the isothermal process $B(t):$ (i)~$B_1 \rightarrow B_2$ and (iii)~$B_3 \rightarrow B_4$ for any form of $B(t)$. To set up the isentropic processes (ii)~$B_2 \rightarrow B_3$ and (iv)~$B_4 \rightarrow B_1$, we must satisfy the condition (see Appendix~\ref{Isoentropic})
\begin{equation}
B_b = \sqrt{ \frac{\beta_a}{\beta_b}\left(B_a^2+E^2 \right) - E^2},
\label{isentropic0}
\end{equation}
where $\beta_a$ and $\beta_b$ are the inverse temperatures of the baths before and after the adiabatic process for $(a, b)= ({\rm ii}, {\rm iii})$ or $({\rm iv}, {\rm i})$; otherwise, entropy would be generated. In the conventional Carnot cycle, the isentropic process is regarded as a temperature-changing process in which the internal energy changes spontaneously under adiabatic conditions. In the present case, the subsystem is isolated and the internal energy does not change without an external field. Moreover, the definition of the temperature during the isentropic process is not clear, because the system is microscopic and isolated except for the initial and final states at $\beta_a$ and $\beta_b$. Thus, instead of assuming a spontaneous change, we actively control the external field (IDF) using Eq.~\eqref{isentropic0} to realize the temperature change between $\beta_a$ and $\beta_b$ in the isentropic processes.

As in the Carnot cycle, the parameters $B_1$ and $B_2$ can take any values, but they must satisfy the relation $B_1 < B_2$ to ensure that the work done by the cycle is positive. In Fig.~\ref{BM0}, we depict the $B$--$M$ diagram for $B_1= \sqrt{15} / 2$ and $B_2= \sqrt{7} / 2$ with $\beta_{\rm 1}/\beta_{\rm 2}=2$. In comparison with the $P$--$V$ diagram for an ideal gas, the $B$--$M$ diagram has the opposite rotational direction. This is because an ideal gas is described by $dU=TdS -PdV$, whereas here we have $dU=T dS +BdM$ (see Eqs. \eqref{dUA}-\eqref{U_B}).  Thus, this cycle is described as (i)~isothermal compression, (ii)~isentropic compression, (iii)~isothermal expansion, and (iv)~isentropic expansion, which is the opposite of the $P$--$V$ case. The area enclosed by the curves corresponds to the work, but a counterclockwise cycle represents positive work, which is also the opposite of the $P$--$V$ case.

\subsection{Cases for nonperturbative and non-Markovian conditions}
\begin{table}
\caption{\label{table:table1}Cycle of the IDF [$B(t)$] and the hot and cold ATFs [$A_1 (t)$ and $A_2(t)$] for an eight-stroke Carnot engine. Because we explicitly treat the bath removal and bath attachment processes of hot and cold baths denoted by (i$'$)--(vi$'$), the cycle consists of eight strokes. 
Here, the stroke intervals are equally spaced and described as $\tau$. Thus, the cycle period is $T=8\tau$. The parameters $B_1$ and $B_2$ can take any values, while $B_3$ and $B_4$ are determined from Eq.~\eqref{isentropic0}. We set $B_1 = \sqrt{15} / 2$, $B_2 = \sqrt{7} / 2$, $B_3 = 1 / 2$, and $B_4 = \sqrt{3} / 2$.\\}
\centering
\begin{tabular}{|c|c|c|c|c|}
\hline
& $B ( t )$ & $A_1 ( t ) / |A_1|$ & $A_2 ( t ) / |A_2|$ \\
\hline \hline
(i) & $B_1 + (B_2 - B_1) t / \tau$ & $1$ & $0$ \\
\hline
(i$'$) & $B_2$ & $1 - ({t - \tau})/{\tau}$ &
$0$ \\
\hline
(ii) & $B_2 + (B_3 - B_2) ({t - 2 \tau})/{\tau}$
& $0$ & $0$ \\
\hline
(ii$'$) & $B_3$ & $0$ & $({t - 3 \tau})/{\tau}$ \\
\hline
(iii) & $B_3 + ( B_4 - B_3 ) (t - 4 \tau)/{\tau}$
& $0$ & $1$ \\
\hline
(iii$'$) & $B_4$ & $0$ & $1 - ({t - 5 \tau})/{\tau}$ \\
\hline
(iv) & $B_4 + ( B_1 - B_4 )({t - 6 \tau})/{\tau}$
& $0$ & $0$ \\
\hline
(iv$'$) & $B_1$ & $({t - 7 \tau})/{\tau}$ & $0$ \\
\hline
\end{tabular}
\end{table}

\begin{figure}[!t]
\begin{center}
\includegraphics[width=7cm]{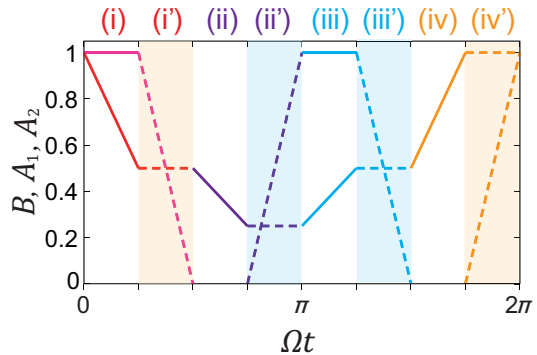}
\end{center}
\caption{Time profiles of $B(t)$, $A_1 (t)$, and $A_2(t)$ for the eight-stroke Carnot cycle depicted as functions of $\Omega t$, where $\Omega = \pi/4 \tau$. The profile of $B(t)$ consists of the red, purple, blue, and orange solid lines, corresponding to isothermal and isentropic processes (i)--(iv), and the red, purple, blue, and orange dashed lines, corresponding to the bath removal and bath attachment processes of hot and cold baths (i$'$)--(vi$'$), respectively. The profiles of $A_1 (t)$ and $A_2(t)$ consist of the orange dashed, red solid, and red dashed lines, and the purple dashed, blue solid, and blue dashed lines, corresponding to the bath attachment, operation, and removal processes (vi$'$), (i), and (i$'$), and (ii$'$), (iii), and (iii$'$), for the hot and cold baths, respectively.
\label{BA1A2}
}
\end{figure}

\subsubsection{Simulation details}
We now simulate the Carnot engine using our model, which includes both the IDF and the ATFs. We compute $W_{\rm tot}(t)=W_{\rm A+I}(t)$ for various cycle frequencies $\Omega$, and then analyze the characteristics of the engine in the quasi-static case using thermodynamic variables defined by the qHE, $\Delta F_{\rm A+I}^{\rm qst}(t) = W_{\rm A+I}^{\rm qst} (t)$. As explained in the factorized case, to realize the isentropic processes, we must change the IDF after removing or before attaching the heat bath, as described by (ii) and (iv). In addition to the regular four Carnot processes (i)--(iv), we must explicitly treat four additional processes that represent (i$'$)~removing and (iv$'$)~attaching the hot bath and (ii$'$)~attaching and (iii$'$)~removing the cold bath. 

We conduct numerical simulations for the TLS coupled to two bosonic baths at different temperatures using the HEOM approach. Here, we consider the equally spaced stroke period $\tau$. Thus, the cycle period and frequency are $T=8 \tau$ and $\Omega = \pi / 4 \tau $, respectively. (See Table~\ref{table:table1} and Fig.~\ref{BA1A2}.) Throughout this paper, we fix the system parameter as $E=0.5$ and the bath parameters as $\beta_1=0.5$, $\beta_2=1.0$, $\gamma \equiv \gamma_1=\gamma_2=1.0$, and $A_0= 0.1, \sqrt{0.1}$, or $\sqrt{3.0}$, where $A_0\equiv A_1=A_2$ is the maximum strength of $A_1 ( t )$ and $A_2 ( t )$. We choose the truncation number of the hierarchy, representing the depth of the HEOM computation, as $N = 8$ for $A_0 = \sqrt{3.0}$ and $N = 6$ for the other cases. A Pad{\'e} spectral decomposition scheme is employed to obtain the expansion coefficients of the noise correlation functions.\cite{hu2010communication} We set the number of Pad{\'e} frequencies to $K_1 = K_2 = 4$. We integrate Eq.~\eqref{ModelHEOM} from the temporal initial state until the simulation cycle reaches the steady state with the time-dependent fields using the fourth-order Runge--Kutta method with a time step of $\delta t=1.0\times 10^{-2}$. The various thermodynamic variables can then be evaluated from the HEOM elements.

To reduce the computation time, we set $\hat{\rho}_{\vec{n}}$ to zero for the elements $N_1 > 1$ and $N_2 > 1$ in $\tau \geq 100$, where $N_k = \sum_{l = 0}^{K_k} n_l^k$. This treatment is valid because the ADOs of the hot bath are negligibly small by the time the bath is restored to the system; the ADOs decay exponentially in time after the baths are removed from the system.

\subsubsection{Thermal efficiency}
\begin{figure}
\includegraphics[width=7cm]{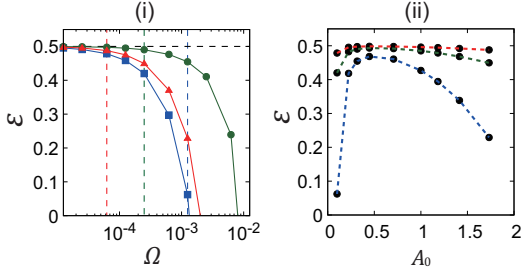}
\caption{\label{TEDiagram}(i) Efficiencies of the quantum Carnot engine calculated as a function of $\Omega$ for different SB coupling strengths: (a)~$A_0= 0.1$ (blue curve with square markers), (b)~$A_0=\sqrt{0.1}$ (green curve with circular markers), and (c)~$A_0 = \sqrt{3.0}$ (red curve with triangular markers). (ii) Efficiencies of the quantum Carnot cycle as a function of $A_0$ at the frequencies $\Omega / 2 \pi = 1.25 \times 10^{-3}$ (blue dashed curve), $2.5 \times 10^{-4}$ (green dashed curve), and $6.25 \times 10^{-5}$ (red dashed curve) marked with black circular data points. The corresponding $\Omega$ is indicated by the dashed line for each frequency in (i).}
\end{figure}

We now investigate the efficiency of the engine, which is significant in the characterization of the Carnot cycle. The total work and heat from the hot bath per cycle, $W_{\rm tot}$ and $Q_{{\rm B}_1}$, are obtained from Eqs.~\eqref{eq:W}--\eqref{qHEWI} and Eqs.~\eqref{expectHB} and \eqref{U_Brd2}. The efficiency is then evaluated as
\begin{align}
\varepsilon= \frac{W_{\rm tot}}{Q_{{\rm B}_1}} .
\label{epsilon}
\end{align}

Figure~\ref{TEDiagram}(i) displays the efficiency $\varepsilon$ as a function of $\Omega$ for (a) the weak ($A_0 = 0.1$), (b) intermediate ($A_0 = \sqrt{0.1}$), and (c) strong ($A_0 = \sqrt{3.0}$) SB coupling cases. For any $A_0$, the efficiency converges to the Carnot limit $\varepsilon_{\rm C} = 1 - \beta_{\rm H} / \beta_{\rm C}$ (which is $0.5$ in the present case) for the quasi-static case $\Omega \rightarrow 0$, providing a manifestation of Carnot's theorem.

Figure~\ref{TEDiagram}(ii) displays the efficiency as a function of the SB coupling strength $A_0$ for the three values of $\Omega$ indicated by the colored dashed lines in Fig.~\ref{TEDiagram}(i). When $A_0$ is small, the thermal excitation is weak and a longer time is required for equilibration. Conversely, when $A_0$ is very large, the efficiency decreases due to the strong relaxation: the heat bath suppresses the kinetic motion relating to the heat flow.\cite{KT15JCP,KT16JCP,ST20JCP} Thus, for a fixed value of $\Omega$, the efficiency reaches a maximum in the intermediate region where $A_0$ is neither large nor small. This feature has been observed in various quantum transport problems, such as exciton transfer\cite{IshizakiFleming2009} and chemical-reaction problems,\cite{TW92JCP} and is known in the classical case as the Kramers turnover problem. As $\Omega$ becomes smaller and approaches the quasi-static limit, the $A_0$ dependence of the efficiency is suppressed and the results approach the Carnot limit, as can be seen from Fig.~\ref{TEDiagram}(i).

\subsubsection{\textit{B}--\textit{M} and \textit{A}--\textit{D} diagrams}

\begin{figure}[!t]
\includegraphics[width=8.2cm]{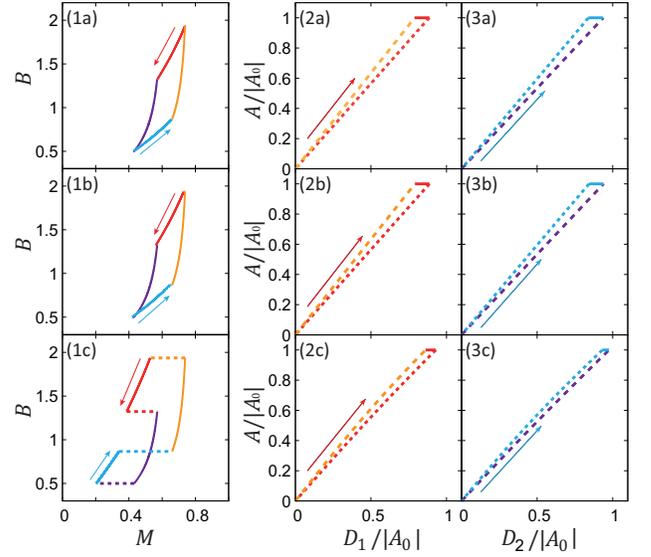}
\caption{\label{BMADDiagram} (1)~$B$--$M$, (2)~$A_1$--$D_1$, and (3)~$A_2$--$D_2$ diagrams for the quasi-static processes with $E=0.5$ in (a) the weak SB coupling case ($A_0= 0.1$), (b) the intermediate SB coupling case ($A_0 = \sqrt{0.1}$), and (c) the strong SB coupling case ($A_0 = \sqrt{3.0}$). In (1), the $B$--$M$ diagrams, the red and blue curves represent the (i)~hot ($\beta_1=0.5$) isothermal compression and (iii)~cold ($\beta_2=1.0$) isothermal expansion processes, whereas the purple and orange curves represent the (ii)~isentropic compression and (iv)~isentropic expansion processes, respectively. The cycle starts from the red arrow and evolves in a counterclockwise fashion over time (heat engine). In (2), the $A_1$--$D_1$ diagrams, the red and orange dashed curves represent (i$'$)~removing and (iv$'$)~attaching the hot bath, whereas in (3), the $A_2$--$D_2$ diagrams, the purple and blue dashed curves represent (ii$'$)~attaching and (iii$'$)~removing the cold bath. The cycles in the $A_1$--$D_1$ and $A_2$--$D_2$ diagrams start from the red and blue arrows and evolve in clockwise and counterclockwise fashions over time (refrigerator and heat engine), respectively.}
\end{figure}

Under the quasi-static condition, we can employ a thermodynamic description of the system because the variables defined from the qHE by Eqs.~\eqref{DefnoneqM} and \eqref{DefnoneqD} become the state variables.\cite{ST20JCP, ST21JPSJ, KT22JCP1} By comparing these thermodynamic variables with the physical quantities calculated by HEOM, we can quantitatively verify the qHE description of quantum thermodynamics.

In Fig.~\ref{BMADDiagram}, we depict the results for the IDF, hot ATF, and cold ATF as the (1)~$B$--$M$, (2)~$A_1$--$D_1$, and (3)~$A_2$--$D_2$ diagrams in the quasi-static case for the (a)~weak, (b)~intermediate, and (c)~strong SB coupling cases. 
In these diagrams, the bath-removal processes (i$'$) and (iii$'$) and bath-attachment processes (ii$'$) and (iv$'$) are represented as dashed curves. The trajectories of the work diagram are periodic and closed because $M$, $D_1$, and $D_2$ are the state variables. The area enclosed by each diagram corresponds to negative work when evolving in a counterclockwise fashion over time, and to positive work when evolving in a clockwise fashion.
We have numerically confirmed that $W_{\rm A+I} (\beta; t)$, evaluated as the areas surrounded by the trajectories of the diagrams, agrees with $W_{\rm A+I}^{\rm HEOM}(t)$ calculated from the HEOM. (See Table \ref{WorkAreaTable}.)

In the case of weak and intermediate SB coupling in Figs.~\ref{BMADDiagram}(2a), \ref{BMADDiagram}(2b), \ref{BMADDiagram}(3a), and \ref{BMADDiagram}(3b), the positive work done by the hot ATF and the negative work done by the cold ATF are approximately equivalent and almost cancel each other out. Thus, the total work is predominantly determined by the work done by the IDF, as illustrated in the $B$--$M$ diagram. In the cases of weak and intermediate coupling, the profiles in Figs.~\ref{BMADDiagram}(1a) and~\ref{BMADDiagram}(1b) thus become similar to that in the factorized assumption case in Fig.~\ref{BA1A2}.

In the cases of strong coupling in Figs.~\ref{BMADDiagram}(1c)--\ref{BMADDiagram}(3c), because the SB interaction includes contributions from both the subsystem side and the bath side, the profiles in the $A$--$D$ and $B$--$M$ diagrams change significantly. If this TLS is regarded as a spin system, $B ( t )$ corresponds to the excitation energy of the spin. Then, as $B ( t )$ increases, the spin becomes aligned with the ground state, so the magnetization $M(\beta; t)$ increases. Because the SB interaction with $\hat{V} = \hat{\sigma}_x$ excites the spin, an increase or decrease in $A_k ( t )$ causes a decrease or increase in $M ( \beta ; t )$, even if $B ( t )$ remains constant. Thus, in the $B$--$M$ diagram in Fig.~\ref{BMADDiagram}(1c), we observe a contribution from the red and orange dashed lines resulting from changes in $A_1(t)$ in processes (i$'$) and (iv$'$) and from the purple and blue dashed lines resulting from changes in $A_2(t)$ in processes (ii$'$) and (iii$'$). The areas surrounded by the trajectories of the $B$--$M$ diagrams are divided vertically into two parts, with the processes evolving in a counterclockwise fashion over time at the top and those evolving in a clockwise fashion over time at the bottom. In this case, the work is evaluated from the area of the top minus the area of the bottom. In Fig.~\ref{BMADDiagram}(1c), such effects appear as the vertical shifts of the solid curves corresponding to (i)--(iv). 

In the case of the $A$--$D$ diagrams, an increase in $B(t)$ suppresses the spin excitation effect of $A_k ( t )$, so $D_k ( \beta ; t )$ decreases even if $A_k ( t )$ does not change, as shown by the red and blue horizontal lines in the $A_1$--$D_1$ and $A_2$--$D_2$ diagrams in Figs.~\ref{BMADDiagram}(2c) and \ref{BMADDiagram}(3c), respectively. The suppression effect of $B(t)$ is smaller for the hot bath than for the cold bath, and the absolute value of the work done is larger for $A_1$ than for $A_2$. The difference between the work done by $A_1$ and that done by $A_2$ is compensated by the work done by $B$, as presented in Table \ref{WorkAreaTable}. As we will show below using a $T$--$S$ diagram, the bath removal and bath attachment processes dominate the heat transfer processes in the strong SB interaction case.

Although the profiles of the $B$--$M$ diagrams are very different when the SB coupling strength is large, the total work done by the external fields is almost the same, regardless of the SB coupling strength, in the quasi-static limit, providing a manifestation of Carnot's theorem. (See Fig. \ref{TEDiagram}(ii) and Table \ref{WorkAreaTable}.)

\begin{widetext}

\begin{table}
\caption{\label{WorkAreaTable}Work done in one cycle for three values of the SB coupling strength $A_0$. 
Here, $W_{a}$ with $a= B$--$M$, $A_1$--$D_1$, and $A_2$--$D_2$ represents the work evaluated as the areas surrounded by the trajectories of the work diagrams in Fig.~
\ref{BMADDiagram}, and $W_{b}^{\rm HEOM}$ with $b=A$, $I_1$, and $I_2$ represents the work directly evaluated from the HEOM approach with Eqs.~\eqref{qHEWA} and \eqref{qHEWI} 
for one cycle of the external fields $B(t)$, $A_1 (t)$, and $A_2 (t)$, respectively. In both cases, the value of the total work is identical and, in the HEOM case, it is evaluated as $W_{\rm tot}^{\rm HEOM}= W_{\rm A}^{\rm HEOM}+ W_{\rm I_1}^{\rm HEOM} + W_{\rm I_2}^{\rm HEOM}$.  Then, as we illustrated in Table \ref{EntropyTable} in Appendix \ref{sec.TimeEntropy}, we have $W_{\rm tot}^{\rm HEOM}+Q_{\rm tot}^{\rm HEOM}=0$\\
}
\begin{tabular}{c|cc|cc|cc|c}
\hline
$A_0$ & $\quad W_{\rm B-M} \quad$ & $\quad W_{\rm A}^{\rm HEOM} \quad $ & $\qquad W_{\rm A_1 - D_1} \qquad$ & $W_{\rm I_1 }^{\rm HEOM}$ & $\qquad W_{\rm A_2 - D_2} \qquad$ & $W_{\rm I_2 }^{\rm HEOM}$ & $\qquad W_{\rm tot}^{\rm HEOM} \qquad$
\\ 
\hline \hline
$0.1$ & $-0.127$ & $-0.127$ & $4.83 \times 10^{-4}$ & $4.83 \times 10^{-4}$ & $-4.90 \times 10^{-4}$ & $-4.90 \times 10^{-4}$ & $-0.127$
\\
\hline
$\sqrt{0.1}$ & $-0.128$ & $-0.128$ & $4.81 \times 10^{-3}$ & $4.81 \times 10^{-3}$ & $-4.85 \times 10^{-3}$ & $-4.85 \times 10^{-3}$ & $-0.129$
\\
\hline
$\sqrt{3.0}$ & $-0.149$ & $-0.149$ & $0.122$ & $0.122$ & $-0.101$ & $-0.101$ & $-0.128$
\\
\hline 
\end{tabular}
\end{table}
\end{widetext}

\subsubsection{\textit{T}--\textit{S} diagram}
\begin{figure}
\centering
\includegraphics[width=3.2cm]{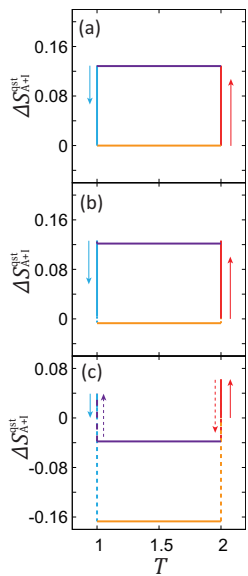}
\caption{\label{TSDiagram} $T$--$S$ work diagrams for the quasi-static process ($\Omega/2 \pi=1.25\times 10^{-5}$) in the (a) weak, (b) intermediate, and (c) strong SB coupling cases, corresponding to (a)--(c) in Fig.~\ref{BMADDiagram}, respectively. The red and blue lines represent the (i) hot isothermal compression and (iii) cold isothermal expansion processes, whereas the purple and orange dashed lines are the (ii) isentropic compression and (iv) isentropic expansion processes, respectively. In each figure, we set $\Delta S_{\rm A+I}^{\rm qst}(0) = 0$ and the cycle starts from the hot isothermal compression, as illustrated by the red arrow. In the strong SB coupling case in (c), the bath removal and attachment processes, represented by the blue and red dashed lines, determine the amount of heat production. }
\end{figure}

We now consider the $T$--$S$ diagram. Here, the change in the quasi-equilibrium Boltzmann entropy (qBE) is calculated from Eq.~\eqref{defqBE} by numerically differentiating $\Delta F^{\rm qst}_{\rm A + I} ( \beta_k ; t )$ with respect to $\beta_k$ for $k=1$ and 2 using the finite difference expression $[\Delta F^{\rm qst}_{\rm A + I} ( \beta_k + \Delta \beta; t ) -\Delta F^{\rm qst}_{\rm A + I} ( \beta_k; t )]/ \Delta \beta$, with $\Delta \beta = 0.0001$.

Although the $B$--$M$ and $A$--$D$ diagrams are useful for interpreting the work that is directly evaluated using the HEOM approach, the situation for the $T$--$S$ diagram is different because the entropy is purely a thermodynamic variable that is indirectly evaluated using the temperature. Therefore, the $T$--$S$ diagram is better suited to exploring the thermodynamic features of the system. The area of the $T$--$S$ diagram is equivalent to the net heat gain, which can be evaluated from the HEOM formalism as the change in the bath energy.
In Fig.~\ref{TSDiagram}, we depict the $T$--$S$ diagram for the (a) weak [$A_0 = 0.1$], (b) intermediate [$A_0 = \sqrt{0.1}$], and (c) strong [$A_0 = \sqrt{3.0}$] SB interaction cases. After the heat bath is removed, the system is isolated and the entropy does not change under the isentropic manipulation described in Sec.~\ref{factorized}. For reference, we also plot the results based on the von Neumann entropy calculated from the zeroth HEOM element as $S_{\rm A}^{\rm vN} ( t ) = - \mathrm{tr}_{\rm A} \{ \hat{\rho}_{\vec{0}} ( t ) \ln \hat{\rho}_{\vec{0}} ( t ) \}$ in Appendix \ref{sec.TimeEntropy}. The trajectories of the $T$--$S$ diagram are periodic and closed because the qBE is the state variable. The area enclosed by each diagram corresponds to the positive heat when evolving in a counterclockwise fashion over time. (See Table \ref{EntropyTable} in Appendix \ref{sec.TimeEntropy}.)

For the weak and intermediate SB coupling cases in Figs.~\ref{TSDiagram}(a) and \ref{TSDiagram}(b), the profiles of the $T$--$S$ diagram are similar to the factorized case in Fig.~\ref{BM0}(b). This is because the contribution of the entropy from the SB interaction is small in these cases. Conversely, for the strong SB coupling case in Fig.~\ref{TSDiagram}(c), the $T$--$S$ diagram differs significantly due to the entropy change associated with the bath removal and bath attachment processes. This is because the system can be excited by the SB interaction through $\hat{V} = \hat{\sigma}_x$, so that entropy increases during the bath attachment process as the heat flowing from the bath to the system increases, whereas entropy decreases during the bath removal process due to reverse heat flow. Thus, although the area enclosed by the lines corresponding to the net heat gain is the same, as illustrated in Table \ref{EntropyTable}, the heat flow is controlled by the bath attachment and removal processes in the strong coupling case, whereas the heat flow is controlled by the isothermal process in the weak and intermediate coupling cases. However, the area enclosed by the trajectory in the $T$--$S$ diagram does not depend on the SB coupling strength, as in the case of work.

\section{\label{sec:conclude}Conclusions}
We have conducted accurate numerical simulations of a quantum Carnot engine based on an SB model, incorporating the ATFs and IDF to describe the manipulations of the adiabatic wall by external forces. The HEOM formalism enables the evaluation of the internal energies of not only the subsystem, but also the bath and the SB interactions, even in low-temperature, non-Markovian, and nonperturbative conditions. We analyzed the real-time responses of work as functions of the cycle frequency. As expected, the computational results approach the Carnot efficiency in the quasi-static limit. Moreover, the efficiency does not change, regardless of the SB coupling strength, which is the manifestation of Carnot's theorem. 

If the ATFs are abruptly changed, as in the conventional theory when introducing an adiabatic process, the cycle is no longer quasi-static and the efficiency becomes lower than the Carnot limit. The use of reduced equations of motion with Markovian and rotating wave approximations, such as the Lindblad equation, should be avoided because this does not accurately treat the quantum entanglement between the subsystem and the bath or the dynamic thermal effects described by the fluctuation and dissipation arising from the heat bath.\cite{T14JCP,T15JCP}

By regarding work as the qHE, we computed the conjugate thermodynamic variables of the IDF and ATFs and depicted the results as thermodynamic work diagrams. We further introduced the qBE in an attempt to characterize the thermal properties of the cycle. We showed that work variables such as $M (\beta; t)$, $D_1(\beta; t)$, and $\Delta S_{{\rm A}+{\rm I}}^{\rm qst} (\beta; t)$ are state variables. Thus, the areas surrounded by the trajectories in the $M$--$B$ and $A$--$D$ diagrams correspond to work, whereas the area within the trajectory of the $T$--$S$ diagram corresponds to the net heat gain. The results show that ATFs are responsible for the work when the SB coupling is very strong, whereas the IDF is responsible when the SB coupling is not strong. As expected, the total work and net heat gain were found to be independent of the SB coupling strength in the quasi-static limit, as predicted by Carnot almost 200 years ago. The inclusion of ATFs is the key to maintaining physical consistency in the study of quantum thermodynamics, and perhaps for quantum information.

Although here we employed a simple spin-boson system, we can apply the same method to ideal gas systems characterized by a $P$--$V$ diagram by using the hierarchical quantum Fokker--Planck equations.\cite{T15JCP,TW92JCP, IT19JCTC,T20JCP} In the case of a bosonic gas, the results should be similar to those for a classical ideal gas, except at very low temperatures; conversely, in the case of a fermionic gas, the results should exhibit quantum effects even at high temperatures. It would be interesting to investigate heat engines with different mechanisms, although it seems unlikely that the thermodynamic laws will be violated.

In this investigation, we limited our analysis to work variables in a quasi-static case. As an extension of this study, it is possible to introduce nonequilibrium free energy using the nonequilibrium work. Such investigations are left for future work.

\section*{Acknowledgments}
Y.T. was supported by JSPS KAKENHI (Grant No.~B21H01884).
S.K. acknowledges a fellowship supported by JST, the establishment of university fellowships towards the creation of science technology innovation (Grant No.~JPMJFS2123).

\section*{Author declarations}
\subsection*{Conflict of Interest}
The authors have no conflicts to disclose.

\section*{Data availability}
The data that support the findings of this study are available from the corresponding author upon reasonable request.

\appendix
\section{Isentropic processes}
\label{Isoentropic}
In the classical Carnot cycle, the isentropic process is regarded as a temperature-changing process in which the internal energy also changes. In the present case, we cannot define the temperature during the isentropic process because the system is microscopic and isolated, aside from the initial and final states described by $\beta_1$ or $\beta_2$. Thus, instead of assuming a spontaneous change in internal energy, we actively control the external field (IDF) to realize the temperature change between $\beta_a$ and $\beta_b$ in the isentropic processes.

Because the heat baths have been removed, we only consider the main system here. For an isentropic process, we have
\begin{equation}
d U_{\rm A} = - M_{\rm A}^0 d B,
\label{dUMA0}
\end{equation}
because the change in entropy is zero. Alternatively, the internal energy is expressed as
\begin{equation}
d U_{\rm A} = \left( \frac{\partial U_{\rm A}}{\partial \beta} \right)_{\rm B} d \beta
+ \left( \frac{\partial U_{\rm A}}{\partial B} \right)_\beta d B.
\label{dUA}
\end{equation}
Note that, in statistical physics, the free energy defined from the partition function is ubiquitously referred to as the Helmholtz energy. In the thermodynamic sense, the Helmholtz energy for $B$ should be called the Gibbs energy because they appear as the Legendre transformation of $F$ with respect to $B$ and its conjugate property $M$ ($dG = -S dT -M dB$).\cite{Oono2017} Accordingly, $dU = TdS +MdB$ should be called enthalpy instead of internal energy. Following convention, however, we refer to $F$ as Helmholtz energy and $U$ as internal energy.

Because the internal energy of the TLS is evaluated from Eq.~\eqref{FA0} as
\begin{equation}
U_{\rm A}^0 (t) = - \sqrt{ B^2(t) + E^2} \tanh ( \beta \sqrt{ B^2(t) + E^2} ),
\label{U_A0}
\end{equation}
we have
\begin{equation}
\left( \frac{\partial U_{\rm A}}{\partial \beta} \right)_{\rm B}
= - \frac{{ B^2(t) + E^2} }{\cosh^2 ( \beta \sqrt{ B^2(t) + E^2} )},
\end{equation}
\begin{align}
\left( \frac{\partial U_{\rm A}}{\partial B} \right)_\beta
= &- \frac{B(t)}{\sqrt{ B^2(t) + E^2} } \tanh ( \beta \sqrt{ B^2(t) + E^2} ) \nonumber \\
-& \frac{\beta B(t)}{\cosh^2 ( \beta\sqrt{ B^2(t) + E^2} )}.
\label{U_B}
\end{align}
From Eqs.~\eqref{dUMA0} and \eqref{dUA}, we have the relation
\begin{align}
- (B^2+ E^2 )d \beta = \beta B d B.
\label{GGG}
\end{align}
For isentropic processes from $B(t_a)=B_a$ at $\beta_a$ to $B(t_b)=B_b$ at $\beta_{\rm B}$, we have
\begin{equation}
- \ln \left( \frac{\beta_b}{\beta_a} \right)
= \ln \left( \frac{\sqrt{E^2 + B_b^2}}{\sqrt{E^2 + B_{a}^2}} \right).
\end{equation}
Thus, we have the relation
\begin{equation}
B_b = \sqrt{ \frac{\beta_a}{\beta_b}\left(B_a^2+E^2 \right) - E^2}.
\label{isentropic}
\end{equation}

\section{Quasi-equilibrium Boltzmann entropy and von Neumann entropy}
\label{sec.TimeEntropy}

\begin{figure}[H]
\centering
\includegraphics[width=5cm]{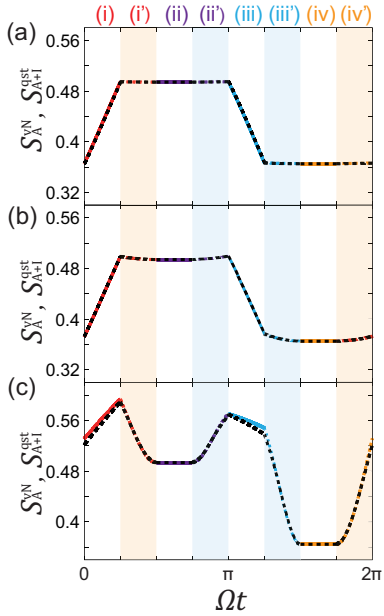}
\caption{\label{TimeSDiagram} Time evolution of the qBE (color curves) and vNE (black dashed curves) in the (a) weak, (b) intermediate, and (c) strong SB coupling cases for the eight-stroke cycle denoted by (i)--(iv'). (For the correspondence between each stroke and the line colors, see Fig.~\ref{BA1A2}.)}
\end{figure}

\begin{table}
\caption{\label{EntropyTable} Net heat gain, evaluated as the area in the $T$--$S$ diagrams based on the qBE [$Q_{qBE}$] and vNE [$Q_{\rm vNE}$] for different SB coupling strengths $A_0$. They are compared with one cycle of the bath energy [$Q_{tot}^{\rm HEOM}\equiv\Delta U_{{\rm B}_1}^{\rm qst} (\beta_1; t)+ \Delta U_{{\rm B}_2}^{\rm qst} (\beta_2; t)$], evaluated using Eq.~\eqref{U_Brd2} from the HEOM approach.
\\}
\begin{tabular}{cccc}
\hline
$A_0$ & $\quad Q_{qBE} \quad$ & $\quad Q_{\rm vNE} \quad$ & $Q_{\rm tot}^{\rm HEOM}$
\\
\hline \hline
$0.1$ & $0.1285$ & $0.1289$ & $0.1272$
\\
\hline
$\sqrt{0.1}$ & $0.1286$ & $0.1289$ & $0.1287$
\\
\hline
$\sqrt{3.0}$ & $0.1292$ & $0.1290$ & $0.1283$
\\
\hline
\end{tabular}
\end{table}

The von Neumann entropy (vNE) is commonly used in quantum thermodynamics, particularly in nonequilibrium cases. It is defined as follows:
\begin{align}
S_A^{\rm vN}(t) = -{\operatorname{tr}_{\rm A}} \{ {\hat \rho_{\rm A} (t)}\ln{\hat \rho_{\rm A} (t)} \},
\label{eq:Neumann-entropyt}
\end{align}
where ${\hat \rho_A (t)}$ is the reduced density matrix. In isolated systems, the qBE and vNE are equivalent, but in a reduced system, this equivalence does not hold due to the contribution of entropy from the SB interaction. If we use the reduced density matrix obtained from nonperturbative approaches, such as the HEOM approach, the description of vNE becomes reasonably accurate, even in the strong SB coupling case.\cite{ST20JCP} Here, using the zeroth member of the HEOM solution in Eq.~\eqref{eq:Neumann-entropyt} as $\hat \rho_A (t)=\hat{\rho}_{\vec{0}}$, we introduce the vNE to characterize the time evolution of the qBE more closely. Because the Carnot cycle involves two isentropic processes, where the qBE and vNE results must agree, we can verify the description of vNE in the isotropic processes, as well as the transitions between the isentropic and isothermal processes.

Using the calculated entropy, we also computed the net heat gain per cycle. For example, for the vNE, $Q_{\rm vNE}$ is evaluated as
\begin{align}
Q_{\rm vNE} = \int_{-\frac{\pi}{4\Omega}}^{\frac{\pi}{2\Omega}} dt T_1 \frac{\partial S_{\rm A}^{\rm vN} (\beta_1; t)}{\partial t}
+ \int_{\frac{3\pi}{4\Omega}}^{\frac{3\pi}{2\Omega}} d t T_2 \frac{\partial S_{\rm A}^{\rm vN} ( \beta_2 ; t )}{\partial t}.
\label{heatproduction}
\end{align}
The calculated results are presented in Table \ref{EntropyTable}. Although the net heat gain calculated directly from HEOM agrees with the total work presented in Table \ref{WorkAreaTable}, there is a slight discrepancy between $Q_{\rm tot}^{\rm HEOM}$ and the other values obtained from the qBE and vNE. This may be due to the insufficient precision of the $\beta$ derivative in evaluating the entropy. 

In Fig.~\ref{TimeSDiagram}, we depict the qBE and vNE as a function of time for different SB coupling strengths. Here, we set the origin of the two graphs as $S_{\rm A+I}^{\rm qst}(t)= S_{\rm A}^{\rm vN}(t)$ at $ t = \pi/2\Omega $. As illustrated in Figs.~\ref{TimeSDiagram}(a) and \ref{TimeSDiagram}(b), the results from the qBE and vNE are similar and almost overlap. This is because the contribution of the entropy from the SB interaction is small in these cases. In the strong coupling case in panel (c), while the time evolution of entropy in the isentropic processes is similar, that in the isothermal processes controlled by $B(t)$ is different. This is because the contribution of the entropy from the SB interaction is not accurately taken into account in the vNE due to its reduced description of the system. Thus, the vNE underestimates the entropy compared with the qBE in the isothermal processes. Because their contribution cancels out, the net heat gain, evaluated as the area in $T$--$S$ diagrams, is similar in both the qBE and vNE cases, as presented in Table \ref{EntropyTable}.

\bibliography{references,tanimura_publist}

\end{document}